\documentclass[a4paper,UKenglish,cleveref,autoref,thm-restate]{lipics-v2021}
\pdfoutput=1 

\usepackage{amsmath}
\usepackage{alltt}
\usepackage{tikz-cd}

\newcommand{\mkhref}[3]{\href{https://github.com/CakeML/game-of-life/blob/itp2025/#1\#L#2}{#3}}

\usepackage{holtexbasic}
\renewcommand{\HOLTokenTurnstile}{\ensuremath{\vdash\!\!}}
\newcommand{\HOLConstNL}[1]{\textsf{\small #1}} 
\renewcommand{\HOLConst}[2][]{\HOLConstNL{#2}}
\renewcommand{\HOLConst}[2][]{\ifthenelse{\equal{#1}{}}{\autolink{#2}{\HOLConstNL{#2}}}{\autolink{#1}{\HOLConstNL{#2}}}}
\renewcommand{\HOLFieldName}[1]{\textsf{\small #1}}
\renewcommand{\HOLSymConst}[1]{\HOLConstNL{#1}}
\renewcommand{\HOLTyOp}[2][]{\HOLConst[#1]{#2}}
\newcommand{\HOLTyOpNL}[1]{\HOLConstNL{#1}} 
\newcommand{\HOLThm}[2][]{\HOLConst[#1]{#2}}

\renewcommand{\HOLKeyword}[1]{\mathsf{#1}}

\renewcommand{\HOLTokenDoublePlus}{\ensuremath{+\rule{-0.2em}{0em}+}}

\renewcommand{\HOLTokenDoublePlus}{\ensuremath{\mathrel{+\mkern-10mu+}}}

\newcommand{\cell}{\HOLConst[Cell]{cell}}
\newcommand{\ck}{\HOLConst[Clock]{ck}}
\newcommand{\this}{\HOLConst[ThisCell]{this}}

\newcommand{\puteqnum}{\refstepcounter{equation}\textup{(\theequation)}}

\newcommand{\fixme}[1]
{ {\color{blue} #1 }}

\makeatletter
\newcommand{\pushright}[1]{\ifmeasuring@#1\else\omit\hfill$\displaystyle#1$\fi\ignorespaces}
\makeatother

\usepackage{xcolor}
\definecolor{red}{HTML}{aa0000}
\definecolor{blue}{HTML}{0000cc}
\definecolor{yellow}{HTML}{aaaa00}
\definecolor{lightyellow}{HTML}{ffff66}
\definecolor{purple}{HTML}{CC00CC}

\usepackage{tcolorbox}

\usepackage{environ}
\NewEnviron{holthmenv}{%
  \[
  \scalebox{1.0}{\begin{array}[t]{l}
  \BODY
  \end{array}}
  \]}

\NewEnviron{holthmenvl}{%
\begin{equation}\begin{aligned}
  \scalebox{1.0}{\begin{array}[t]{l}\BODY\end{array}}
\end{aligned}\end{equation}}

\newcommand{\yellowbox}[1]{%
  \tcbox[boxrule=0pt,on line,boxsep=1.3pt,left=0pt,right=0pt,top=0pt,bottom=0pt,colback=lightyellow]{#1}}

\usepackage{proof}

\hideLIPIcs  

\title{GOL in GOL in HOL}
\subtitle{Verified Circuits in Conway's Game of Life}

\titlerunning{GOL in GOL in HOL}

\author{Magnus O. Myreen}{Chalmers University of Technology and University of Gothenburg, Gothenburg, Sweden}{myreen@chalmers.se}{https://orcid.org/0000-0002-9504-4107}{}
\author{Mario Carneiro}{Chalmers University of Technology and University of Gothenburg, Gothenburg, Sweden} {marioc@chalmers.se}{https://orcid.org/0000-0002-0470-5249}{}

\Copyright{Magnus O. Myreen and Mario Carneiro}

\authorrunning{M. O. Myreen and M. Carneiro}



\funding{This work was supported by the Swedish Research Council (grant no. 2021-05165)}

\nolinenumbers 

\acknowledgements{
  A special thank you to Andreas L\"o\"ow for sowing the seeds for this work by pointing the first author to the blog posts by Nicholas Carlini.
}

\begin{document}

\newcommand{\autolink}[2]{
\IfEqCase{#1}{%
{A}{\mkhref{gol_simScript.sml}{45}{#2}}%
{adj}{\mkhref{gol_rulesScript.sml}{16}{#2}}%
{B}{\mkhref{gol_simScript.sml}{45}{#2}}%
{bexp}{\mkhref{gol_simScript.sml}{50}{#2}}%
{build_mega_cells}{\mkhref{gol_in_gol_circuitScript.sml}{16}{#2}}%
{Cell}{\mkhref{gol_simScript.sml}{468}{#2}}%
{circ_mod}{\mkhref{gol_circuitScript.sml}{213}{#2}}%
{circ_mod_wf}{\mkhref{gol_circuitScript.sml}{75}{#2}}%
{circuit_run}{\mkhref{gol_circuitScript.sml}{344}{#2}}%
{Clock}{\mkhref{gol_in_gol_circuit2Script.sml}{39}{#2}}%
{E}{\mkhref{gol_simScript.sml}{10}{#2}}%
{floodfill}{\mkhref{gol_in_gol_circuit2Script.sml}{1217}{#2}}%
{floodfill_run}{\mkhref{gol_in_gol_circuit2Script.sml}{929}{#2}}%
{infl}{\mkhref{gol_lemmasScript.sml}{88}{#2}}%
{io_step}{\mkhref{gol_io_stepScript.sml}{27}{#2}}%
{io_steps}{\mkhref{gol_io_stepScript.sml}{47}{#2}}%
{is_gate}{\mkhref{gol_in_gol_circuit2Script.sml}{315}{#2}}%
{live_adj}{\mkhref{gol_rulesScript.sml}{20}{#2}}%
{N}{\mkhref{gol_simScript.sml}{10}{#2}}%
{nextCell}{\mkhref{gol_in_gol_circuit2Script.sml}{104}{#2}}%
{read_mega_cells}{\mkhref{gol_in_gol_circuit2Script.sml}{1372}{#2}}%
{run}{\mkhref{gol_io_stepScript.sml}{57}{#2}}%
{S}{\mkhref{gol_simScript.sml}{10}{#2}}%
{step}{\mkhref{gol_rulesScript.sml}{25}{#2}}%
{ThisCell}{\mkhref{gol_in_gol_circuit2Script.sml}{39}{#2}}%
{W}{\mkhref{gol_simScript.sml}{10}{#2}}%
{avalue}{\mkhref{gol_in_gol_circuit2Script.sml}{37}{#2}}%
{bexp}{\mkhref{gol_simScript.sml}{49}{#2}}%
{dir}{\mkhref{gol_simScript.sml}{10}{#2}}%
{evalue}{\mkhref{gol_in_gol_circuit2Script.sml}{39}{#2}}%
{gate}{\mkhref{gol_in_gol_circuit2Script.sml}{297}{#2}}%
{modifier}{\mkhref{gol_io_stepScript.sml}{19}{#2}}%
{stream}{\mkhref{gol_in_gol_circuit2Script.sml}{74}{#2}}%
{value}{\mkhref{gol_in_gol_circuit2Script.sml}{40}{#2}}%
{var}{\mkhref{gol_simScript.sml}{45}{#2}}%
{floodfill_add_crossover_l}{\mkhref{gol_in_gol_circuit2Script.sml}{2118}{#2}}%
{floodfill_add_gate}{\mkhref{gol_in_gol_circuit2Script.sml}{1992}{#2}}%
{floodfill_add_ins}{\mkhref{gol_in_gol_circuit2Script.sml}{1246}{#2}}%
{floodfill_finish_crossover}{\mkhref{gol_in_gol_circuit2Script.sml}{2178}{#2}}%
{floodfill_teleport}{\mkhref{gol_in_gol_circuit2Script.sml}{2239}{#2}}%
{gol_in_gol_circuit_thm}{\mkhref{gol_in_gol_circuitScript.sml}{28}{#2}}%
}[\fixme{missing link:} #2]}

\maketitle

\begin{abstract}

\noindent
Conway's Game of Life (GOL) is a cellular automaton that has captured the interest of hobbyists and mathematicians alike for more than 50 years. The Game of Life is Turing complete, and people have been building increasingly sophisticated constructions within GOL, such as 8-bit displays, Turing machines, and even an implementation of GOL itself.  In this paper, we report on a project to build an implementation of GOL within GOL, via logic circuits, fully formally verified within the HOL4 theorem prover. This required a combination of interactive tactic proving, symbolic simulation, and semi-automated forward proof to assemble the components into an infinite circuit which can calculate the next step of the simulation while respecting signal propagation delays. The result is a verified ``GOL in GOL compiler'' which takes an initial GOL state and returns a mega-cell version of it that can be passed to off-the-shelf GOL simulators, such as Golly. We believe these techniques are also applicable to other cellular automata, as well as for hardware verification which takes into account both the physical configuration of components and wire delays.

\end{abstract}

\maketitle



\section{Introduction}
\label{sec:intro}

Conway's Game of Life is a cellular automaton that was first published in the Scientific American in 1970 where it was said to model the rise and fall of societies~\cite{gardner1970gameoflife}.
Since its initial publication, GOL has remained a curiosity among hobbyists and mathematicians for its combination of simplicity and surprisingly chaotic organic look as a model of computation.
It has been demonstrated that one can create order in the chaos and build interesting constructions in GOL such as, e.g., simple computers, complete Turing machines, or even simulations of GOL in GOL~\cite{gameoflife,Todesco2013}.

In this paper, we build infrastructure for formal reasoning about circuits in GOL and construct a verified circuit in GOL that can simulate GOL itself. However, before we describe our work, we provide necessary background on GOL.

\subsection{A short introduction to Conway's Game of Life}

Conway's Game of Life is a deterministic simulation that is performed on an unbounded two-dimensional grid of cells. Each cell can be either alive or dead. Time passes in discrete steps and, at each step, all cells simultaneously update to their next state. The state of a cell at location $(i,j)$ at time $n+1$ is determined by its state at time $n$ and its immediate neighbours' states at time $n$.
Cell $(i,j)$ is live at time $n+1$, if and only if:
\begin{itemize}
\item cell $(i,j)$ is live at time $n$ and two or three of its neighbours are live at time $n$, or
\item cell $(i,j)$ is dead at time $n$ and exactly three of its neighbours are live at time $n$.
\end{itemize}
The neighbours of a cell $(i,j)$ are the eight cells that are adjacent to it, e.g., $(i,j+1)$, or share a corner, e.g., $(i+1,j+1)$.

The rules of the game do not restrict the initial state, i.e., the state at time 0.
The challenge is to find initial states that lead to interesting behaviour when the simulation is run.
There are numerous GOL simulators\footnote{For example: \url{https://playgameoflife.com/}} with which people can experiment with different initial configurations of the GOL grid.
Anyone who has tried drawing a busy initial pattern in a GOL simulator will have observed that GOL quickly evolves into a chaotic mess that often looks like digital depiction of the evolution of a bacteria culture.

\subsubsection*{Gliders and spaceships}

In this world of GOL chaos, there are however certain patterns that are well behaved and can be used in interesting ways. The simplest kind of pattern, called a ``still life,'' is pattern which stays unchanged on the next clock cycle. These patterns will simply remain static until something else interacts with them. There are also oscillators which go through a sequence of states before returning to the original state.

\begin{figure}[tb]
\centering
\includegraphics[width=\linewidth]{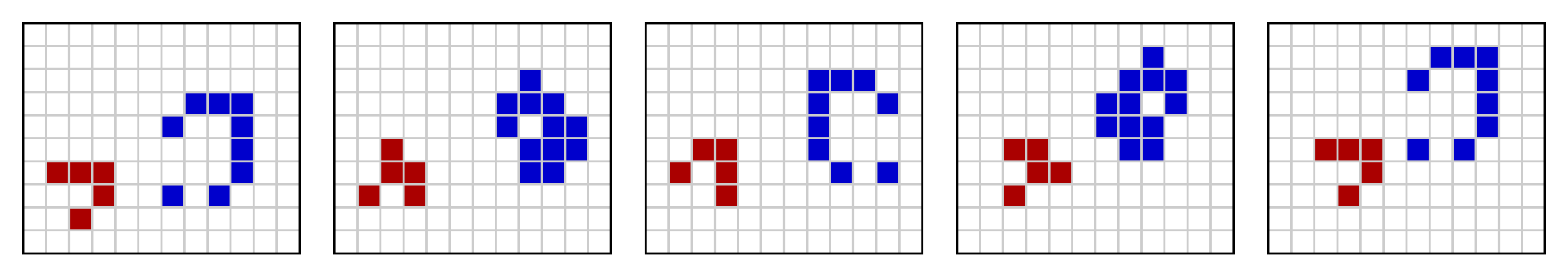}
\caption{Evolution of the glider ({\color{red}red}, left) and the lightweight spaceship (LWSS) ({\color{blue}blue}, right).}\label{fig:spaceships}
\end{figure}

A slightly more interesting class of patterns are called ``spaceships,'' which are similar to oscillators but with a spatial shift. The most famous spaceship is called the \emph{glider}. If left undisturbed, it will in four time steps transform itself into exactly its own original shape but shifted one step diagonally. Over time, gliders move across the grid diagonally, at a speed of $1/4$ cells per clock cycle. Another spaceship is the \emph{lightweight spaceship} (LWSS), which move horizontally or vertically at a speed of $1/2$ cells per clock cycle. Figure~\ref{fig:spaceships} shows how the glider and LWSS move across the grid.

\subsubsection*{Useful behaviour through collisions}
\begin{figure}[tb]
\centering
\includegraphics[width=\linewidth]{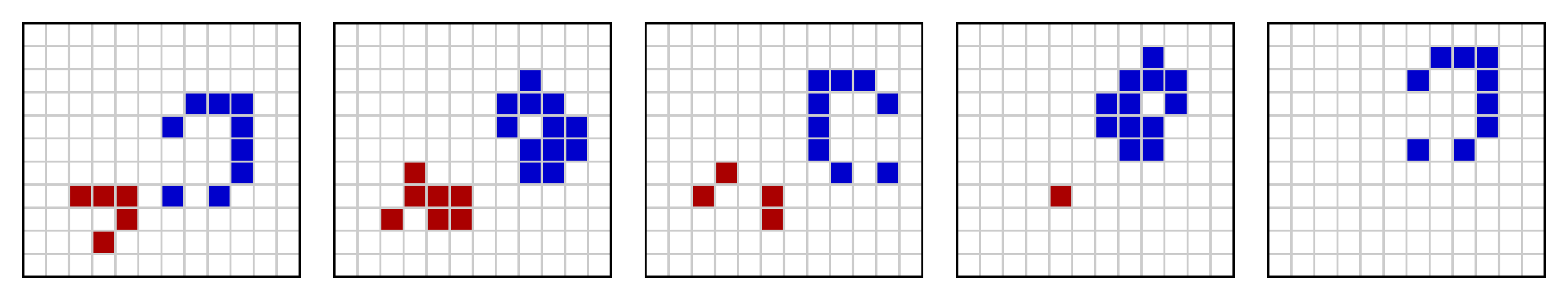}
\caption{Collision between a glider and an LWSS which destroys the glider but leaves the LWSS intact.}\label{fig:collision}
\end{figure}

While gliders and spaceships are cute on their own, their real use comes into focus when we observe that they can carry signals and these signals can be processed via well behaved collisions of (streams of) gliders and spaceships. Collisions can have a variety of effects, including destroying one or both of the incoming spaceships, producing other spaceships, maybe in different directions, and making a big mess of random figures (we will be deliberately avoiding this kind of collision). Figure~\ref{fig:collision} shows a collision between a glider and a LWSS which destroys the glider, but the LWSS is unimpeded.

Because gliders are so small, it is not so difficult for them to be spontaneously created, and a key component of our design (and indeed most GOL constructions) utilizes the \emph{Gosper glider gun}, a pattern that produces a glider every 30 ticks. This is what enables us to have steady streams to collide in the first place.

To get things larger than gliders, one can use gliders as a fusion reaction component. Not all kinds of spaceships can be produced in this way, but the LWSS can be produced by three gliders colliding in just the right way, and we will be using this to produce LWSS streams.

If we view a spaceship stream as carrying a timed sequence of bits with a $1$ where a ship is present and a $0$ when one is absent, then a spaceship stream $(a_i)$ which is intercepted by another stream $(b_i)$ which cleanly destroys both can be seen as computing the logical operation $(a_i\wedge \neg b_i)$, because the output at position $i$ is present only if ship $a_i$ was present, and it was not knocked out by ship $b_i$. (It also computes $b_i\wedge\neg a_i$ for the other stream, but usually we will only be interested in one output.)

Empty space gives us the constant sequence $0$, and the glider gun or its variations give us the sequence $1$, so we can already see how we can get a NOT gate as $1\wedge \neg a$, an AND gate using $a\wedge\neg(1\wedge\neg b)$, and so we seem to have a complete system of logic gates already. The devil is in the details, but this basic intuition is broadly correct and the rest is engineering.

\begin{figure}[tbp]
  \centering
  \begin{subfigure}[b]{0.475\textwidth}
    \centering
    \includegraphics[width=\textwidth]{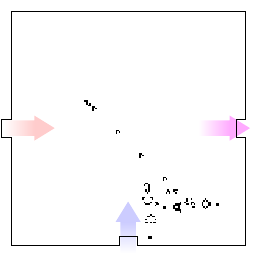}%
    \vspace{-3pt}
    \caption{\small AND gate computing ${\color{red}0}\wedge{\color{blue}0}={\color{purple}0}$}
    \label{fig:and-00}
  \end{subfigure}
  \hfill
  \begin{subfigure}[b]{0.475\textwidth}
    \centering
    \includegraphics[width=\textwidth]{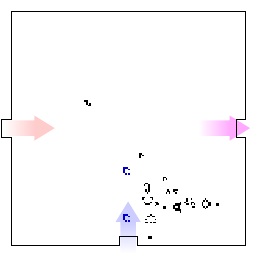}%
    \vspace{-3pt}
    \caption{\small AND gate computing ${\color{red}0}\wedge{\color{blue}1}={\color{purple}0}$}
    \label{fig:and-01}
  \end{subfigure}
  \begin{subfigure}[b]{0.475\textwidth}
    \centering
    \includegraphics[width=\textwidth]{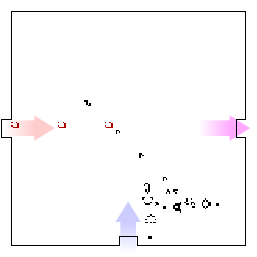}%
    \vspace{-3pt}
    \caption{\small AND gate computing ${\color{red}1}\wedge{\color{blue}0}={\color{purple}0}$}
    \label{fig:and-10}
  \end{subfigure}
  \hfill
  \begin{subfigure}[b]{0.475\textwidth}
    \centering
    \includegraphics[width=\textwidth]{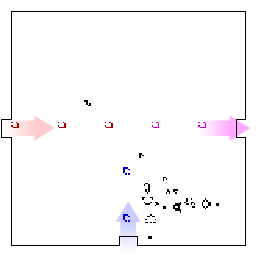}%
    \vspace{-3pt}
    \caption{\small AND gate computing ${\color{red}1}\wedge{\color{blue}1}={\color{purple}1}$}
    \label{fig:and-11}
  \end{subfigure}
  \begin{subfigure}[b]{0.475\textwidth}
    \centering
    \includegraphics[width=\textwidth]{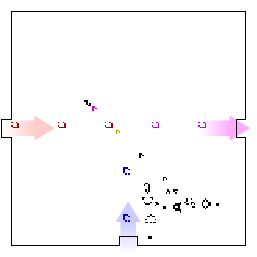}%
    \vspace{-3pt}
    \caption{\small AND gate, symbolic simulation of ${\color{red}a}\wedge{\color{blue}b}$ (\S\ref{sec:gate-sim})}
    \label{fig:and}
  \end{subfigure}
  \hfill
  \begin{subfigure}[b]{0.475\textwidth}
    \centering
    \includegraphics[width=\textwidth]{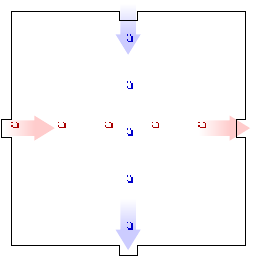}%
    \vspace{-3pt}
    \caption{\small A crossover gate.}
    \label{fig:cross}
  \end{subfigure}
  \caption{\small Illustration of some gates and their data flow behavior. Color key: ${\color{red}a}, {\color{blue}b}, {\color{yellow}\neg b}, {\color{purple}a\wedge b}$.}
  \label{fig:gates}
\end{figure}

Nicholas Carlini (and others, e.g., \cite{Todesco2013}) have demonstrated that one can build digital gates based on such collisions. In our work, we use Carlini's gates and his circuit conventions as a starting point. Figure~\ref{fig:gates} shows how one of his AND gates works. This particular gate takes in streams of LWSSs travelling {\color{red}east} and {\color{blue}north}, and essentially computes AND using the formula ${\color{red}a}\wedge\neg(\neg {\color{blue}b}\wedge 1)$ as we described.

An additional aspect of Carlini's design is that the streams have period 60 instead of period 30. In Figure~\ref{fig:gates} one can see a second glider gun which takes out every other glider produced from the first gun so that the $1$ stream has period 60. The reason for this convention is so that LWSS streams can pass through each other without collision (Figure~\ref{fig:cross}), which gives us a ``crossover gate'', a key tool for building logic circuits in the plane.

In this paper, we describe how one can formally verified circuits in GOL, and in particular, how we have constructed a verified GOL circuit that implements GOL itself.
The work we presented here has been carried out in the HOL4 theorem prover~\cite{HOL4}.
The final product is a verified ``GOL in GOL compiler'' that, given a GOL configuration, produces a tiling of logic gates implemented in GOL such that the overall behaviour of the circuit is a simulation of GOL starting from the given GOL configuration.

\section{Mega-Cell and Approach}
\label{sec:approach}

\begin{figure}[tbp]
\centering
\includegraphics[width=\linewidth]{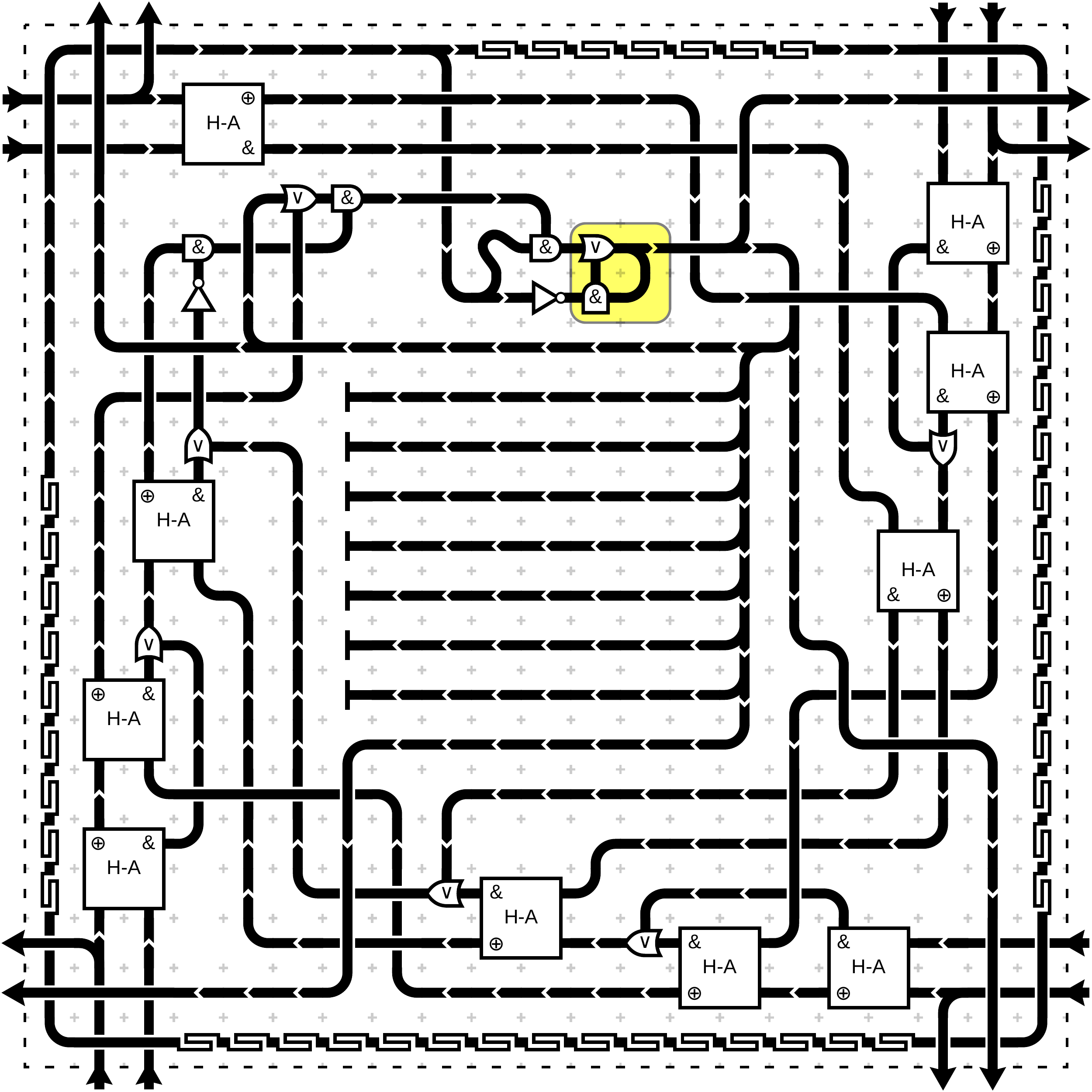}
\caption{A high-level circuit diagram representation of the mega-cell in our construction. There are AND (\textsf{\&}), OR ($\vee$) and NOT ($\rhd\!\circ$) gates, half-adders (H-A), and all of it is acyclic except for the latch, highlighted in \yellowbox{yellow}, and the clock, which is a wire cycle forming the outer border of the cell. The clock uses slow wires (visualized as switchbacks) in order to ensure that the main logic can complete in time for the next clock cycle.}\label{fig:mega-cell}
\end{figure}

The technical work for this project had as its goal to build a verified mega-cell as shown in Figure~\ref{fig:mega-cell}. The purpose of this mega-cell is to implement one GOL cell in a space completely tiled with copies of the mega-cell. As the entire grid is tiled with mega-cells performing similar computations, this verified circuit built inside GOL simulates GOL itself. Each of the gates and wires in the diagram are implemented by gates similar to Carlini's gates.

The circuit in Figure~\ref{fig:mega-cell} works as follows: at the heart of each mega-cell is a latch, i.e., the little loop highlighted in \yellowbox{yellow}. This latch holds the state of the GOL cell this mega-cell is simulating. The output of this latch is lead through wires to all neighbouring cells, as can be seen by following the wires from the latch. Because all neighbouring cells have the same circuit wiring, whenever it gives away its value to a neighbor in one of the outgoing arrows from the cell, there is a corresponding input from the opposite side where the neighbor shares its value to this cell, and ultimately the 8 incoming arrows provide the values of all 8 surrounding cells. These inputs are then lead through half-adders (H-A) to sum up the number of neighbours that are alive at the moment.

While this computation is being performed, there is also a train of LWSSs cycling slowly on the big loop around the perimeter of the mega-cell. This train of LWSSs is the clock, and it is timed so that just as the value is finished computing, the clock pulse reaches the latch input, and the latch updates its state to the next value this cell is supposed to represent. The twisty wires make the clock signal go slowly enough to not tick the latch forward before the next value is ready.

The mega-cell is laid out on a grid with components which are $1\times 1$ or $2\times 2$ tiles large, and each tile in this grid is $150\times 150$ GOL cells (Figure~\ref{fig:gates} shows a some of these tiles). The mega-cell is $21\times 21$ tiles large, meaning that the full construction is $3150 \times 3150$ GOL cells in size. Proving the correctness of such a large object requires careful design of abstraction layers in order to be manageable.

Our approach consists of several layers of abstraction. The work is organised into the following high-level steps:
\begin{enumerate}

\item We formalise the rules of GOL, define the notion of a GOL pattern's area of influence, and show that two patterns evolve indpendently if their areas of influence are disjoint.

\item In order to locally prove properties of patterns that communicate with adjacent patterns, we define GOL-IO, an alternative semantics for GOL that allows for input and output. We prove composition and input-output internalisation theorems for these.

\item We then formalise the what it means for a pattern to implement a gate, such as Carlini's gates. This involves defining exactly how streams of inputs and result in streams of outputs. Again, we prove composition theorems and input-output internalisation.

\item Next, we define the notion of signals that only carry meaningful values some of the time. This is important since variations in wire delay cause signals to ``jitter'' for a while before settling down to the correct value.

\item We then compose all of the gates in our mega-cell together to obtain a proof that it calculates a specific logic formula, and does so within the time budget provided by the clock signal.

\item Finally, we prove that the formula that is computed is the GOL step function, and that therefore a complete tiling of the space with mega-cells results in the desired GOL simulation.
\end{enumerate}

While many theorems were proved using traditional goal-oriented tactic proofs, some of the
heavy lifting was done by automation, including in-logic computation.
\begin{enumerate}

\item We proved high-level specifications for individual gates using symbolic simulation of the GOL rules.

\item We also automated the composition of all of the gate specifications that make up a mega-cell. This automation computes how and when each wire will have a specific value.

\end{enumerate}
Both of these, but particularly the former, made significant use of HOL4's recently added
feature for fast kernel computation~\cite{cvcompute}.


\section{Formally reasoning about GOL}
\label{sec:gol}

This section describes how we formalise the rules of GOL and our approach to modular verification of patterns in GOL.

\subsection{Rules of GOL}\label{sec:gol-rules}

We define a GOL state as a set $S\subseteq\mathbb{Z}^2$, where $(i,j) \in S$ means that $(i,j)$ is alive in state $S$. As the definition of GOL's next-state function depends on counting the number of live cells neighboring a cell, we use the \HOLConst{live_adj} function to count the number of live neighbours:
\begin{holthmenv}
  \HOLConst{adj}\,\HOLFreeVar{i}\,\HOLFreeVar{j}\,\HOLTokenDefEquality{}\{(i',j')\mid \HOLConstNL{max}\,(|i'-i|,|j'-j|)=1\}\\[0.2em]
  \HOLConst{live_adj}\;\HOLFreeVar{S}\;\HOLFreeVar{i}\;\HOLFreeVar{j}\;\HOLTokenDefEquality{}\;\HOLConstNL{card}\;(\HOLFreeVar{S}\;\HOLSymConst{\HOLTokenInter{}}\;\HOLConst{adj}\;\HOLFreeVar{i}\;\HOLFreeVar{j})
\end{holthmenv}

Now, given a state $S$, $(i,j)$ will be alive in the next state if its number of live neighbors is 2 or 3 if $(i,j)$ is live, or exactly 3 if $(i,j)$ is dead.
\begin{holthmenv}
  \HOLConst{step}\;\HOLFreeVar{S}\;\HOLTokenDefEquality{}\;\{\;(\HOLFreeVar{i}\HOLSymConst{,}\HOLFreeVar{j})\;|\;
\HOLKeyword{if}\;(\HOLFreeVar{i}\HOLSymConst{,}\HOLFreeVar{j})\;\HOLSymConst{\HOLTokenIn{}}\;\HOLFreeVar{S}\;\HOLKeyword{then}\;\HOLConst{live_adj}\;\HOLFreeVar{S}\;\HOLFreeVar{i}\;\HOLFreeVar{j}\in\{2,3\}\;\HOLKeyword{else}\;\HOLConst{live_adj}\;\HOLFreeVar{S}\;\HOLFreeVar{i}\;\HOLFreeVar{j}\;\HOLSymConst{=}\;\HOLNumLit{3} \}
\end{holthmenv}

\subsection{Area of influence and compositionality}

In order to enable modular reasoning about patterns in GOL, we need some notion of non-interference.
The intuition we follow is that two patterns in GOL will not interfere with one another as long as they are sufficiently far from each other.

We formalise this intuition by defining a function \HOLSymConst{infl} which computes the \emph{area of influence} of a GOL state. Location $(i,j)$ is in the area of influence of the patterns in GOL state $S$ if it is at most one step away from a live cell in $S$.
\begin{holthmenv}
  \HOLConst{infl}\,\HOLFreeVar{S}\,\HOLTokenDefEquality{}\,\{\,(i',j')\mid \exists\,i\,j.\,(i,j) \in S\,{\land}\,\HOLConstNL{max}\,(|i'-i|,|j'-j|) \leq 1\,\}
\end{holthmenv}

Using \HOLSymConst{infl} we can capture the intuition that, if two patterns $s$ and $t$ are sufficiently far from each other, then they evolve independently of each other in the next step.
``Sufficiently far'' can be asserted by simply requiring that their areas of influence are disjoint.
\begin{holthmenv}
  \HOLTokenTurnstile{}\;\HOLConst{infl}\;\HOLFreeVar{S}\;\HOLSymConst{\HOLTokenInter{}}\;\HOLConst{infl}\;\HOLFreeVar{S'}\;\HOLSymConst{=}\;\HOLSymConst{\HOLTokenEmpty{}}\;\HOLSymConst{\HOLTokenImp{}}\;\HOLConst{step}\;(\HOLFreeVar{S}\;\HOLSymConst{\HOLTokenUnion{}}\;\HOLFreeVar{S'})\;\HOLSymConst{=}\;\HOLConst{step}\;\HOLFreeVar{S}\;\HOLSymConst{\HOLTokenUnion{}}\;\HOLConst{step}\;\HOLFreeVar{S'}
\end{holthmenv}

Note that, while $\HOLConst{infl}\;\HOLFreeVar{S}\;\HOLSymConst{\HOLTokenInter{}}\;\HOLConst{infl}\;\HOLFreeVar{S'}\;\HOLSymConst{=}\;\HOLSymConst{\HOLTokenEmpty{}}$ is a sufficient condition, it is not the weakest condition one can have on that theorem.
We chose to use this simple condition because it makes proofs easier than more fine-grained conditions.

\subsection{GOL-IO}\label{sec:gol-io}

The patterns in GOL that we set out to verify communicate by sending spaceship patterns to one another. For example, the AND gates shown in Figure~\ref{fig:gates} receive input from the left and the bottom, and produce output to the right. The input and output consist of LWSSs that they receive or hand over to surrounding patterns in GOL.

In order to verify each gate in isolation, we must therefore sever the links between the gates and replace them by an interface. Because every gate will only interact with the interfaces of its neighbors, the precise details of evolution inside the gate will not matter. The way we express this is through a modified step relation called \HOLConst{io_step}:
\begin{holthmenv}
  \HOLConst{io_step}\;\HOLFreeVar{c}\;\ensuremath{\HOLFreeVar{S}\sb{\mathrm{1}}}\;\ensuremath{\HOLFreeVar{S}\sb{\mathrm{3}}}\;\HOLTokenDefEquality{}\\
\;\;\HOLSymConst{\HOLTokenExists{}}\ensuremath{\HOLBoundVar{S}\sb{\mathrm{2}}}.\\
\;\;\;\;\HOLConst{infl}\;\ensuremath{\HOLFreeVar{S}\sb{\mathrm{1}}}\;\HOLSymConst{\HOLTokenSubset{}}\;\HOLFreeVar{c}.\HOLFieldName{area}\;\HOLSymConst{\HOLTokenConj{}}\;\HOLConst{step}\;\ensuremath{\HOLFreeVar{S}\sb{\mathrm{1}}}\;\HOLSymConst{=}\;\ensuremath{\HOLBoundVar{S}\sb{\mathrm{2}}}\;\HOLSymConst{\HOLTokenConj{}}\\
\;\;\;\;\ensuremath{\HOLBoundVar{S}\sb{\mathrm{2}}}\;\HOLSymConst{\HOLTokenInter{}}\;\HOLFreeVar{c}.\HOLFieldName{assert_area}\;\HOLSymConst{=}\;\HOLFreeVar{c}.\HOLFieldName{assert_content}\;\HOLSymConst{\HOLTokenConj{}}\\
\;\;\;\;\ensuremath{\HOLFreeVar{S}\sb{\mathrm{3}}}\;\HOLSymConst{=}\;\HOLFreeVar{c}.\HOLFieldName{insertions}\;\HOLSymConst{\HOLTokenUnion{}}\;(\ensuremath{\HOLBoundVar{S}\sb{\mathrm{2}}}\;\HOLSymConst{\ensuremath{-}}\;\HOLFreeVar{c}.\HOLFieldName{deletions})
\end{holthmenv}

This relation is functional, but unlike \HOLConst{step} it is not total. It is parameterized by a ``modifier'' $c$ which does several things at once:
\begin{itemize}
  \item $\HOLFreeVar{c}.\HOLFieldName{area}$ provides ``guard rails'' for the simulation. The initial state must stay within $\HOLFreeVar{c}.\HOLFieldName{area}$ and must not touch the interior border.
  \item $\HOLFreeVar{c}.\HOLFieldName{assert_area}$ and $\HOLFreeVar{c}.\HOLFieldName{assert_content}$ allow the modifier to assert that a particular pattern appears in the simulation on this step, without otherwise modifying the behavior.
  \item $\HOLFreeVar{c}.\HOLFieldName{insertions}$ and $\HOLFreeVar{c}.\HOLFieldName{deletions}$ actually change the state.
  \begin{itemize}
    \item Inputs can be placed on the board at any time using $\HOLFreeVar{c}.\HOLFieldName{insertions}$.
    \item Outputs are cleanly zapped from the state using $\HOLFreeVar{c}.\HOLFieldName{deletions}$.
  \end{itemize}
\end{itemize}


For most steps,
$\HOLFreeVar{c}.\HOLFieldName{insertions}$,
$\HOLFreeVar{c}.\HOLFieldName{deletions}$,
$\HOLFreeVar{c}.\HOLFieldName{assert_area}$ and
$\HOLFreeVar{c}.\HOLFieldName{assert_content}$ are all empty.
However, at time points when an input is supposed to arrive,
$\HOLFreeVar{c}.\HOLFieldName{insertions}$ will contain an LWSS at an input port.
Similarly, output is handled by a combination of
$\HOLFreeVar{c}.\HOLFieldName{deletions}$,
$\HOLFreeVar{c}.\HOLFieldName{assert_area}$ and
$\HOLFreeVar{c}.\HOLFieldName{assert_content}$ --- $\HOLFreeVar{c}.\HOLFieldName{assert_area}$ and
$\HOLFreeVar{c}.\HOLFieldName{assert_content}$ ensure that the expected output was produced, and $\HOLFreeVar{c}.\HOLFieldName{deletions}$ removes the output from our local simulation.

An important feature of \HOLConst{io_step} is that matching inputs and outputs cancel out. That is, if $\HOLFreeVar{c}.\HOLFieldName{assert_content} = \HOLFreeVar{c}.\HOLFieldName{insertions}$ and $\HOLFreeVar{c}.\HOLFieldName{assert_area} = \HOLFreeVar{c}.\HOLFieldName{deletions}$, then $\HOLConst{io_step}\;\HOLFreeVar{c}\;\ensuremath{\HOLFreeVar{S}\sb{\mathrm{1}}}\;\ensuremath{\HOLFreeVar{S}\sb{\mathrm{3}}}$ implies $\HOLConst{step}\;S_1 = S_3$, i.e., that the insertions and deletions have no effect. This will be relevant later, for the composition theorem.

\subsection{GOL-IO runs}

To reason about runs consisting of many GOL-IO steps, we define $\HOLConst{io_steps}\;k$ which performs $k$-steps of \HOLConst{io_step}. Since \HOLConst{io_step} requires a modifier \HOLFreeVar{c}, \HOLConst{io_steps} requires a sequence of modifiers $c:\mathbb{N}\to\HOLTyOp{modifier}$.
\begin{holthmenv}
  \HOLConst{io_steps}\;\HOLNumLit{0}\;\HOLFreeVar{c}\;\HOLFreeVar{n}\;\ensuremath{\HOLFreeVar{S}\sb{\mathrm{1}}}\;\ensuremath{\HOLFreeVar{S}\sb{\mathrm{2}}}\;\HOLTokenDefEquality{}\;\ensuremath{\HOLFreeVar{S}\sb{\mathrm{1}}}\;\HOLSymConst{=}\;\ensuremath{\HOLFreeVar{S}\sb{\mathrm{2}}}\\
\HOLConst{io_steps}\;(\HOLConstNL{Suc}\;\HOLFreeVar{k})\;\HOLFreeVar{c}\;\HOLFreeVar{n}\;\ensuremath{\HOLFreeVar{S}\sb{\mathrm{1}}}\;\ensuremath{\HOLFreeVar{S}\sb{\mathrm{3}}}\;\HOLTokenDefEquality{}\;\HOLSymConst{\HOLTokenExists{}}\ensuremath{\HOLBoundVar{S}\sb{\mathrm{2}}}.\;\HOLConst{io_step}\;(\HOLFreeVar{c}\;\HOLFreeVar{n})\;\ensuremath{\HOLFreeVar{S}\sb{\mathrm{1}}}\;\ensuremath{\HOLBoundVar{S}\sb{\mathrm{2}}}\;\HOLSymConst{\HOLTokenConj{}}\;\HOLConst{io_steps}\;\HOLFreeVar{k}\;\HOLFreeVar{c}\;(\HOLFreeVar{n}\;\HOLSymConst{\ensuremath{+}}\;\HOLNumLit{1})\;\ensuremath{\HOLBoundVar{S}\sb{\mathrm{2}}}\;\ensuremath{\HOLFreeVar{S}\sb{\mathrm{3}}}\\[0.75em]
\HOLConst{run}\;\HOLFreeVar{c}\;\HOLFreeVar{S}\;\HOLTokenDefEquality{}\;\HOLSymConst{\HOLTokenForall{}}\HOLBoundVar{k}.\;\HOLSymConst{\HOLTokenExists{}}\HOLBoundVar{S'}.\;\HOLConst{io_steps}\;\HOLBoundVar{k}\;\HOLFreeVar{c}\;0\;\HOLFreeVar{S}\;\HOLBoundVar{S'}
\end{holthmenv}

The $\HOLConst{run}\;c\;S$ function asserts that the execution starting from state $S$ is able to run indefinitely using and respecting the modifiers in $c$. Note that the assertions in $c$ can be used to assert that desired values appear at points of interest in the simulation. It is through these assertions we record the behaviour of the verified circuits we build.

\section{Verified Circuits in GOL}
\label{sec:circuits}

This section describes how we write specifications for logic gates built within GOL, the key theorems for working with them, and the abstractions we build to reach the level where we can construct the verified circuit that implements GOL itself.

\subsection{GOL-IO runs for circuit components}
\label{sec:circuit-run}

We state our circuit specifications in terms of \HOLConst{circuit_run} which is defined
in terms of \HOLConst{run} from the previous section.
The meaning of the parameters, \HOLFreeVar{area}, \HOLFreeVar{ins}, \HOLFreeVar{outs}, \HOLFreeVar{init}, and
helper functions, \HOLConst{circ_mod} and \HOLConst{circ_mod_wf},
will be explained later.
\begin{holthmenv}
  \HOLConst{circuit_run}\;\HOLFreeVar{area}\;\HOLFreeVar{ins}\;\HOLFreeVar{outs}\;\HOLFreeVar{init}\;\HOLTokenDefEquality{}\\
\;\;\HOLConst{run}\;(\HOLConst{circ_mod}\;\HOLFreeVar{area}\;\HOLFreeVar{ins}\;\HOLFreeVar{outs})\;\HOLFreeVar{init}\;\HOLSymConst{\HOLTokenConj{}}\\
\;\;\HOLConst{circ_mod_wf}\;\HOLFreeVar{area}\;\HOLFreeVar{ins}\;\HOLFreeVar{outs}
\end{holthmenv}

To get a sense of what circuit specifications look like using an example, consider the AND gate from Figure~\ref{fig:gates}. We can prove the following specification theorem:
\begin{holthmenv}
  \HOLTokenTurnstile{}\;\HOLConst{circuit_run}\ \{(0,0)\}\ \{((-1,0),\HOLConst{E},{\color{red}a}),((0,1),\HOLConst{N},{\color{blue}b})\}\ \{((1,0),\HOLConst{E},{\color{red}a}^{[5]}\wedge{\color{blue}b}^{[6]}\}\ \HOLConstNL{and_gate_pattern}
\end{holthmenv}

The coordinates and delays here are in a higher level coordinate system, where 1 unit corresponds to 75 GOL cells (or 1/2 of a tile) and 1 tick of delay corresponds to 60 GOL steps. The components of the specification are as follows:
\begin{itemize}
\item $area:=\{(0,0)\}$ asserts that this circuit uses one full tile centered at position $(0,0)$. (Because the coordinate system here is in half-tile units, an adjacent gate would be at $(2,0)$. Gates are always placed at double-even coordinates.)
\item $ins:=\{((-1,0),\HOLConst{E},{\color{red}a}),((0,1),\HOLConst{N},{\color{blue}b})\}$ states that this gate has two inputs. The first one is at position $(-1,0)$ (the left edge of the tile), moving east (\HOLConst{E}) into the tile, and carrying some signal $({\color{red}a_t})$. Note that signals are functions from natural numbers to booleans, where ${\color{red}a_t}$ is the value that arrives into this circuit on tick $t$. The second input comes from $(0,1)$, moves north (\HOLConst{N}), and is carrying signal $({\color{blue}b_t})$.
\item $outs := \{((1,0),\HOLConst{E},{\color{red}a}^{[5]}\wedge{\color{blue}b}^{[6]}\}$ states that there is one output stream appearing at location $(1,0)$, moving east out of the tile and carrying signal ${\color{red}a}^{[5]}\wedge{\color{blue}b}^{[6]}$, where $(a^{[n]})_t\HOLTokenDefEquality{}t\ge n\wedge a_{t-n}$ delays a signal by $n$ ticks, and $(a\wedge b)_t\HOLTokenDefEquality{}(a_t\wedge b_t)$ is pointwise AND on signals.
\item $init := \HOLConstNL{and_gate_pattern}$ specifies that the initial configuration of the GOL cells of this AND gate is the content of \HOLConstNL{and_gate_pattern}.
\end{itemize}

\subsection{Input and output in GOL circuits}
\label{sec:circuitio}

As can be seen in Figure~\ref{fig:gates}, circuit tiles have a square geometry with little ports on the sides through which communication happens. We will now discuss how the IO ports are handled. A gate simulation involves the following stages:
\begin{enumerate}
  \item The initial state is set up, as in Figure~\ref{fig:gates}.
  \item IO ports are included for \HOLConst{E}/\HOLConst{W} inputs and \HOLConst{N}/\HOLConst{S} outputs, and excluded for \HOLConst{N}/\HOLConst{S} inputs and \HOLConst{E}/\HOLConst{W} outputs. (See Figure~\ref{fig:cross}.)
  \item 30 GOL steps are performed, during which nothing must escape the bounds.
  \item Deletions are performed at \HOLConst{N}/\HOLConst{S} output ports.
  \item Insertions are performed at \HOLConst{N}/\HOLConst{S} input ports.
  \item IO ports are included for \HOLConst{N}/\HOLConst{S} inputs and \HOLConst{E}/\HOLConst{W} outputs, and excluded for \HOLConst{E}/\HOLConst{W} inputs and \HOLConst{N}/\HOLConst{S} outputs.
  \item 30 GOL steps are performed, during which nothing must escape the bounds.
  \item Deletions are performed at \HOLConst{E}/\HOLConst{W} output ports.
  \item Insertions are performed at \HOLConst{E}/\HOLConst{W} input ports.
  \item Steps 2-9 are repeated for each tick.\footnote{For reasons we will get into in section~\ref{sec:gate-sim}, because of our use of symbolic evaluation we only actually need to perform steps 2-9 once, but the described gate evolution repeats these steps on each tick.}
\end{enumerate}
These steps are all expressible through a carefully chosen sequence of GOL-IO modifiers (see Section~\ref{sec:gol-io}).

In other words, for ports going \HOLConst{N}/\HOLConst{S}, the IO action scheduled to happen on this tick happens at the end of 30 GOL steps (halfway through the tick), while for \HOLConst{E}/\HOLConst{W} ports the action happens after all 60 GOL steps (right at the end of the tick period, before the start of the next tick). The IO action itself is a handoff of an LWSS (or not, depending on the value of the high level signal on that time step) at each output port, and a receipt of an LWSS (or not) at each input port.

The reason for the flip-flopping IO port ownership in steps 2 and 6 is because the producer gate must have ownership of the region prior to the handoff in order to get an LWSS to migrate to that position, and once the deletions and insertions of steps 4,5 are performed, the region must be given to the consumer so the LWSS can get out of the port area and into the consumer gate. The phase difference between \HOLConst{N}/\HOLConst{S} and \HOLConst{E}/\HOLConst{W} ports is to enable crossovers as demonstrated in Figure~\ref{fig:cross}.

\subsection{Composing circuit tiles}\label{sec:circuit-comp}

Equipped with \HOLConst{circuit_run} and an understanding for how input-output ports work, we now look at how \HOLConst{circuit_run} specifications can be composed.

We use the following theorem when composing two \HOLConst{circuit_run} specifications. The theorem requires the areas owned by these specifications to be disjoint. Furthermore, input (resp.\ output) port at the edge of one circuit to have a matching output (resp.\ input) port in the other circuit. Here we overload notation: $(0,0) + \textsf{E} = (1,0)$, and $(0,0) - \textsf{E} = (-1,0)$.
\begin{holthmenv}
  \HOLTokenTurnstile{}\;\HOLConst{circuit_run}\;\ensuremath{\HOLFreeVar{a}\sb{\mathrm{1}}}\;\ensuremath{\HOLFreeVar{ins}\sb{\mathrm{1}}}\;\ensuremath{\HOLFreeVar{outs}\sb{\mathrm{1}}}\;\ensuremath{\HOLFreeVar{init}\sb{\mathrm{1}}}\;\HOLSymConst{\HOLTokenConj{}}\;\HOLConst{circuit_run}\;\ensuremath{\HOLFreeVar{a}\sb{\mathrm{2}}}\;\ensuremath{\HOLFreeVar{ins}\sb{\mathrm{2}}}\;\ensuremath{\HOLFreeVar{outs}\sb{\mathrm{2}}}\;\ensuremath{\HOLFreeVar{init}\sb{\mathrm{2}}}\;\HOLSymConst{\HOLTokenConj{}}\;\ensuremath{\HOLFreeVar{a}\sb{\mathrm{1}}} \cap\ensuremath{\HOLFreeVar{a}\sb{\mathrm{2}}} = \emptyset\;\HOLSymConst{\HOLTokenConj{}}\\
\;\;\;(\HOLSymConst{\HOLTokenForall{}}\HOLBoundVar{p}\;\HOLBoundVar{d}\;\HOLBoundVar{r}.\\
\;\;\;\;\;\;((\HOLBoundVar{p}\HOLSymConst{,}\HOLBoundVar{d}\HOLSymConst{,}\HOLBoundVar{r})\;\HOLSymConst{\HOLTokenIn{}}\;\ensuremath{\HOLFreeVar{ins}\sb{\mathrm{1}}}\;\HOLSymConst{\HOLTokenConj{}}\;\HOLBoundVar{p} - d\;\HOLSymConst{\HOLTokenIn{}}\;\ensuremath{\HOLFreeVar{a}\sb{\mathrm{2}}}\;\HOLSymConst{\HOLTokenImp{}}\;(\HOLBoundVar{p}\HOLSymConst{,}\HOLBoundVar{d}\HOLSymConst{,}\HOLBoundVar{r})\;\HOLSymConst{\HOLTokenIn{}}\;\ensuremath{\HOLFreeVar{outs}\sb{\mathrm{2}}})\;\HOLSymConst{\HOLTokenConj{}}\\
\;\;\;\;\;\;((\HOLBoundVar{p}\HOLSymConst{,}\HOLBoundVar{d}\HOLSymConst{,}\HOLBoundVar{r})\;\HOLSymConst{\HOLTokenIn{}}\;\ensuremath{\HOLFreeVar{outs}\sb{\mathrm{1}}}\;\HOLSymConst{\HOLTokenConj{}}\;\HOLBoundVar{p} + d\;\HOLSymConst{\HOLTokenIn{}}\;\ensuremath{\HOLFreeVar{a}\sb{\mathrm{2}}}\;\HOLSymConst{\HOLTokenImp{}}\;(\HOLBoundVar{p}\HOLSymConst{,}\HOLBoundVar{d}\HOLSymConst{,}\HOLBoundVar{r})\;\HOLSymConst{\HOLTokenIn{}}\;\ensuremath{\HOLFreeVar{ins}\sb{\mathrm{2}}})\;\HOLSymConst{\HOLTokenConj{}}\\
\;\;\;\;\;\;((\HOLBoundVar{p}\HOLSymConst{,}\HOLBoundVar{d}\HOLSymConst{,}\HOLBoundVar{r})\;\HOLSymConst{\HOLTokenIn{}}\;\ensuremath{\HOLFreeVar{ins}\sb{\mathrm{2}}}\;\HOLSymConst{\HOLTokenConj{}}\;\HOLBoundVar{p} - \HOLBoundVar{d}\;\HOLSymConst{\HOLTokenIn{}}\;\ensuremath{\HOLFreeVar{a}\sb{\mathrm{1}}}\;\HOLSymConst{\HOLTokenImp{}}\;(\HOLBoundVar{p}\HOLSymConst{,}\HOLBoundVar{d}\HOLSymConst{,}\HOLBoundVar{r})\;\HOLSymConst{\HOLTokenIn{}}\;\ensuremath{\HOLFreeVar{outs}\sb{\mathrm{1}}})\;\HOLSymConst{\HOLTokenConj{}}\\
\;\;\;\;\;\;((\HOLBoundVar{p}\HOLSymConst{,}\HOLBoundVar{d}\HOLSymConst{,}\HOLBoundVar{r})\;\HOLSymConst{\HOLTokenIn{}}\;\ensuremath{\HOLFreeVar{outs}\sb{\mathrm{2}}}\;\HOLSymConst{\HOLTokenConj{}}\;\HOLBoundVar{p} + \HOLBoundVar{d}\;\HOLSymConst{\HOLTokenIn{}}\;\ensuremath{\HOLFreeVar{a}\sb{\mathrm{1}}}\;\HOLSymConst{\HOLTokenImp{}}\;(\HOLBoundVar{p}\HOLSymConst{,}\HOLBoundVar{d}\HOLSymConst{,}\HOLBoundVar{r})\;\HOLSymConst{\HOLTokenIn{}}\;\ensuremath{\HOLFreeVar{ins}\sb{\mathrm{1}}}))\;\HOLSymConst{\HOLTokenImp{}}\\
\;\;\;\;\;\HOLConst{circuit_run}\;(\ensuremath{\HOLFreeVar{a}\sb{\mathrm{1}}}\;\HOLSymConst{\HOLTokenUnion{}}\;\ensuremath{\HOLFreeVar{a}\sb{\mathrm{2}}})\;(\ensuremath{\HOLFreeVar{ins}\sb{\mathrm{1}}}\;\HOLSymConst{\HOLTokenUnion{}}\;\ensuremath{\HOLFreeVar{ins}\sb{\mathrm{2}}})\;(\ensuremath{\HOLFreeVar{outs}\sb{\mathrm{1}}}\;\HOLSymConst{\HOLTokenUnion{}}\;\ensuremath{\HOLFreeVar{outs}\sb{\mathrm{2}}})\;(\ensuremath{\HOLFreeVar{init}\sb{\mathrm{1}}}\;\HOLSymConst{\HOLTokenUnion{}}\;\ensuremath{\HOLFreeVar{init}\sb{\mathrm{2}}})
\end{holthmenv}

\begin{figure}[t]
  \centering
  \begin{subfigure}[b]{0.328\textwidth}
    \centering
    \caption{}\label{fig:comp1}
    \vspace{-7pt}
    \makebox[\textwidth][c]{\includegraphics[width=1.08\textwidth]{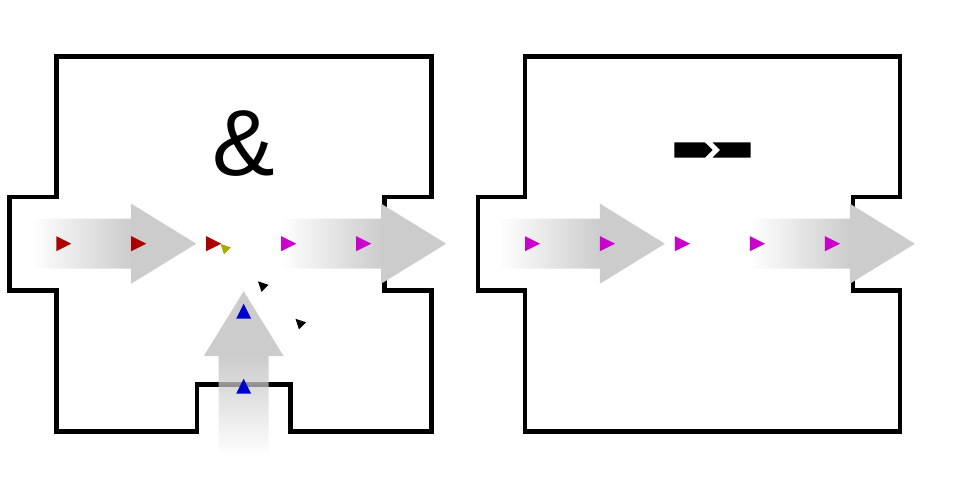}}%
  \end{subfigure}
  \begin{subfigure}[b]{0.328\textwidth}
    \centering
    \caption{}\label{fig:comp2}
    \vspace{-7pt}
    \makebox[\textwidth][c]{\includegraphics[width=1.08\textwidth]{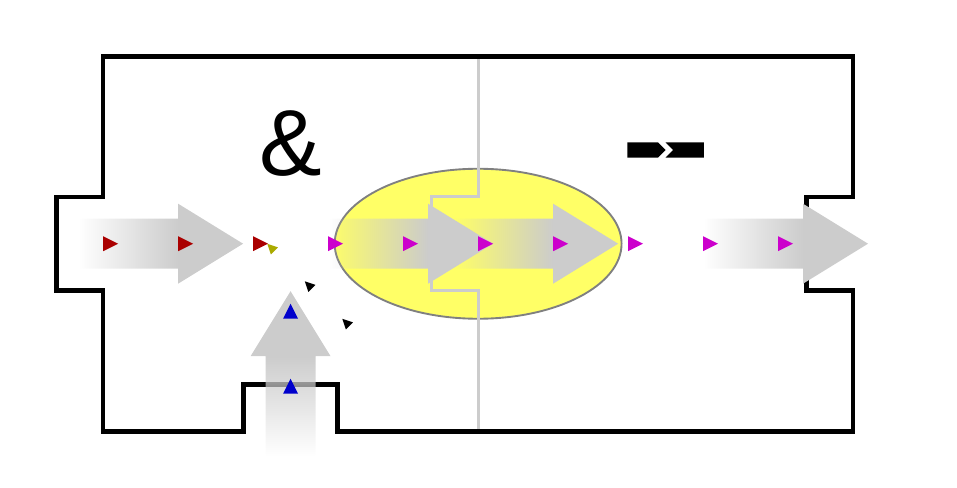}}%
  \end{subfigure}
  \begin{subfigure}[b]{0.328\textwidth}
    \centering
    \caption{}\label{fig:comp3}
    \vspace{-7pt}
    \makebox[\textwidth][c]{\includegraphics[width=1.08\textwidth]{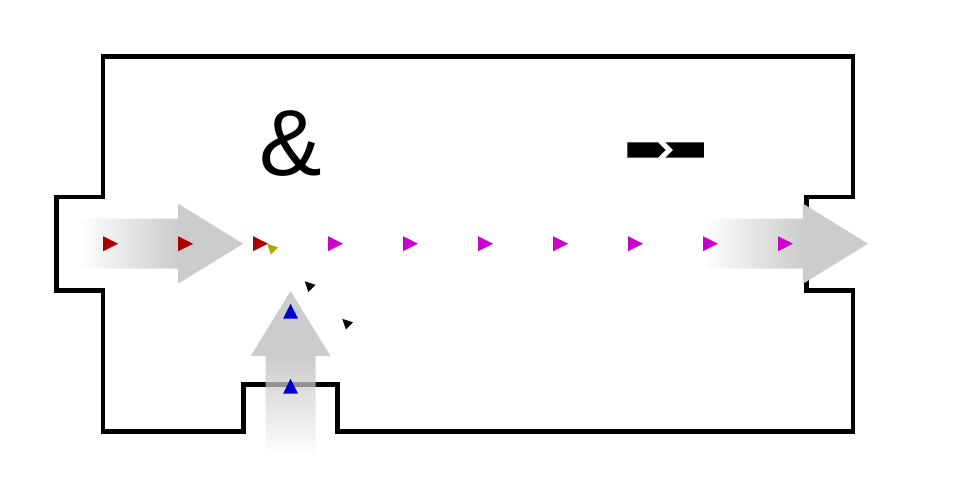}}%
  \end{subfigure}
  \caption{\small Composition of an AND gate and a wire. In step $(a)\to(b)$ we merge the gate areas, resulting in an assembly with an internal input port overlapping an output, highlighted in \yellowbox{yellow}. In step $(b)\to(c)$ the matching pair is canceled.}
  \label{fig:comp}
\end{figure}

The conclusion of the composition theorem above unions each component of the two circuits. This means that matching input and output ports then appear both as input and output ports of the resulting \HOLConst{circuit_run} specification (Figure~\ref{fig:comp2}). The following input-output internalization theorem allows us to delete matching IO ports.
\begin{holthmenv}
  \HOLTokenTurnstile{}\;\HOLConst{circuit_run}\;\HOLFreeVar{area}\;\HOLFreeVar{ins}\;\HOLFreeVar{outs}\;\HOLFreeVar{init}\;\HOLSymConst{\HOLTokenConj{}}\;\HOLFreeVar{m}\;\HOLSymConst{\HOLTokenSubset{}}\;\HOLFreeVar{ins}\;\HOLSymConst{\HOLTokenConj{}}\;\HOLFreeVar{m}\;\HOLSymConst{\HOLTokenSubset{}}\;\HOLFreeVar{outs}\;\HOLSymConst{\HOLTokenImp{}}\\
\;\;\;\;\;\HOLConst{circuit_run}\;\HOLFreeVar{area}\;(\HOLFreeVar{ins}\;\HOLSymConst{\ensuremath{-}}\;\HOLFreeVar{m})\;(\HOLFreeVar{outs}\;\HOLSymConst{\ensuremath{-}}\;\HOLFreeVar{m})\;\HOLFreeVar{init}
\end{holthmenv}

To illustrate, suppose we want to compose an AND gate and a wire, as depicted in figure~\ref{fig:comp}.

\begin{holthmenv}
\rlap{$
1.\ \HOLTokenTurnstile{}\;\HOLConst{circuit_run}\ \{(0,0)\}\ \{((-1,0),\HOLConst{E},{\color{red}a}),((0,1),\HOLConst{N},{\color{blue}b})\}
$}{\hspace{28em}\mbox{AND gate spec}}\\
~~~~~~~~\{((1,0),\HOLConst{E},{\color{red}a}^{[5]}\wedge{\color{blue}b}^{[6]}\}\ \HOLConstNL{and_gate_pattern}\\[0.3em]
\rlap{$%
2.\ \HOLTokenTurnstile{}\;\HOLConst{circuit_run}\ \{(0,0)\}\ \{((-1,0),\HOLConst{E},{\color{red}a})\}\ \{((1,0),\HOLConst{E},{\color{red} a}^{[5]}\}\ \emptyset
$}{\hspace{28em}\mbox{wire spec}}\\[0.3em]
\rlap{$%
3.\ \HOLTokenTurnstile{}\;\HOLConst{circuit_run}\ \{(2,0)\}\ \{((1,0),\HOLConst{E},{\color{red}a})\}\ \{((3,0),\HOLConst{E},{\color{red} a}^{[5]}\}\ \emptyset
$}{\hspace{28em}\mbox{translate (2)}}\\[0.3em]
\rlap{$%
4.\ \HOLTokenTurnstile{}\;\HOLConst{circuit_run}\ \{(2,0)\}\ \{((1,0),\HOLConst{E},{\color{red}a}^{[5]}\wedge{\color{blue}b}^{[6]})\}
$}{\hspace{28em}\mbox{substitute (3)}}\\
~~~~~~~~\{((3,0),\HOLConst{E},({\color{red}a}^{[5]}\wedge{\color{blue}b}^{[6]})^{[5]}\}\ \emptyset\\[0.3em]
\rlap{$%
5.\ \HOLTokenTurnstile{}\;\HOLConst{circuit_run}\ \{(0,0),(2,0)\}
$}{\hspace{28em}\mbox{compose (1,4)}}\\
~~~~~~~~\{((-1,0),\HOLConst{E},{\color{red}a}),((0,1),\HOLConst{N},{\color{blue}b}),((1,0),\HOLConst{E},{\color{red}a}^{[5]}\wedge{\color{blue}b}^{[6]})\}\\
~~~~~~~~\{((1,0),\HOLConst{E},{\color{red}a}^{[5]}\wedge{\color{blue}b}^{[6]}),((3,0),\HOLConst{E},({\color{red}a}^{[5]}\wedge{\color{blue}b}^{[6]})^{[5]}\}\\
~~~~~~~~\HOLConstNL{and_gate_pattern}\\[0.3em]
\rlap{$%
6.\ \HOLTokenTurnstile{}\;\HOLConst{circuit_run}\ \{(0,0),(2,0)\}\ \{((-1,0),\HOLConst{E},{\color{red}a}),((0,1),\HOLConst{N},{\color{blue}b})\}
$}{\hspace{28em}\mbox{internalize (5)}}\\
~~~~~~~~\ \{((3,0),\HOLConst{E},{\color{red}a}^{[10]}\wedge{\color{blue}b}^{[11]}\}\ \HOLConstNL{and_gate_pattern}
\end{holthmenv}

The AND gate specification is familiar from section~\ref{sec:circuit-run}. The wire is similar, but it does not need any initial pattern because LWSSs can travel through empty space. We first translate the wire by $(2,0)$ so it lies next to the AND gate, then substitute ${\color{red}a}$ to ${\color{red}a}^{[5]}\wedge{\color{blue}b}^{[6]}$ so that it matches with the output of the AND gate. We can then compose them in step 5, and the redundant input/output pair is cancelled in step 6, with the delay distributing into the expression.

Here we used a binary version of the composition theorem but in our formalization we also prove and use a more general form of the composition theorem which can compose an arbitrary set of circuits. This is used in particular to tile $\mathbb{Z}^2$-many copies of a gate.

\subsection{Approximate signals}\label{sec:approximate}

The above-described composition process results in exact descriptions of the stream outputs from a set of gates. However, we run into issues when taking into account delay mismatches where the same signal travels via two paths, resulting in expressions like ${\color{red}a}^{[5]}\wedge {\color{red}a}^{[6]}$ which we cannot simplify to ${\color{red}a}^{[6]}$, even though all we really care about is that the ${\color{red}a}$ signal arrives in at most $n$ ticks (in this case, $n=6$).

In fact, this issue can even arise within a single gate. The half-adder is supposed to calculate the XOR of two signals on one output and AND on the other output. However, the actual circuit theorem we obtain looks like this:
\begin{holthmenv}
  \HOLTokenTurnstile{}\;\HOLConst{circuit_run}\ \{(0,0),(2,0),(0,2),(2,2)\}\ \{((-1,0),\HOLConst{E},{\color{red}a}),((-1,2),\HOLConst{E},{\color{blue}b})\}\\
~~~~~\{((3,0),\HOLConst{E},
  ({\color{red}a}^{[15]}\wedge
    (({\color{red}a}^{[12]}\wedge\neg {\color{blue}b}^{[18]})\vee
      (\neg {\color{red}a}^{[12]}\wedge{\color{blue}b}^{[15]}\wedge\neg {\color{blue}b}^{[18]})))\vee(\neg {\color{red}a}^{[15]}\wedge({\color{red}a}^{[12]}\vee{\color{blue}b}^{[15]}))),\\
~~~~~~~((3,2),\HOLConst{E},{\color{red}a}^{[17]}\wedge{\color{blue}b}^{[15]})\}\\
~~~~~\HOLConstNL{half_adder_gate_pattern}
\end{holthmenv}

If we could erase all the delays from the expression at $(3,0)$, we would be able to simplify it to simply ${\color{red}a}^{[?]}\oplus{\color{blue}b}^{[?]}$, but because the delays are different it is simply a somewhat arbitrary function on four boolean values ${\color{red}a}^{[12]},{\color{red}a}^{[15]},{\color{blue}b}^{[15]},{\color{blue}b}^{[18]}$. To make matters worse, this self-delay issue compounds as we push through more gates --- if we were to feed this expression into another XOR we would get even more mirror copies of the signals.

To resolve this, we weaken our constraints on signals. Rather than specifically asserting that a signal is equal to a given value, each wire is associated to an element in a type \HOLTyOp{value}, whose denotation is a \emph{set} of possible signals. This amounts to a Hoare-triple-like precondition at each ``program point'' (= IO port).

We use $s\vdash A$ to assert that $s:\HOLTyOp{stream}$ is in the denotation of $A:\HOLTyOp{value}$, where $\HOLTyOp{stream}\HOLTokenDefEquality\mathbb{Z}^2\to\mathbb{N}\to\HOLTyOpNL{bool}$. A stream has values $s_z(t)$ saying whether there is an LWSS or not each tick, parameterized over $z\in\mathbb{Z}^2$, which is the mega-cell index.

There are two main kinds of signals, \emph{exact signals} and \emph{approximate signals}. Most wires in the circuit carry approximate signals.
\begin{holthmenv}
  \HOLTyOp{avalue}=\cell(\mathbb{Z},\mathbb{Z})\mid \neg \HOLTyOp{avalue}\mid \HOLTyOp{avalue}\wedge \HOLTyOp{avalue}\mid \HOLTyOp{avalue}\vee \HOLTyOp{avalue}\mid \HOLTyOp{avalue}\oplus \HOLTyOp{avalue}\\
  \HOLTyOp{evalue}=\ck\mid \neg \ck\mid \this\mid \this\wedge \ck \mid \this\wedge \neg\ck\\
  \HOLTyOp{value}=\HOLTyOp{avalue}^{[\mathbb{N}]}\mid \HOLTyOp{evalue}^{[\mathbb{Z}]}\mid \top
\end{holthmenv}

An approximate signal ${\color{red}A}^{[n]}$ represents a signal that holds the value ${\color{red}A}:\HOLTyOp{avalue}$ after $n$ ticks, and can have any value before that point. Signals have a refinement relation $A\subseteq B\ \HOLTokenDefEquality\ \forall s.\;s\vdash A\Rightarrow s\vdash B$, and we have ${m\le n\implies {\color{red}A}^{[m]}\subseteq {\color{red}A}^{[n]}}$.

Exact signals are much rarer, and they deal with all of the signals that are involved in the latch and clock, where it is important that we do not allow garbage values to enter. In order for exact signals to support a lossless ${(-)}^{[n]}$ delay operation, an exact signal denotes a $\mathbb{Z}\to\HOLTyOpNL{bool}$ stream, that is, one that extends into the past (even though our GOL simulation normally only deals with $\mathbb{N}\to\HOLTyOpNL{bool}$ streams). There are two main exact signals, which can be combined in a limited way by logical operators:
\begin{itemize}
  \item \ck, the clock signal, is 1 from tick 0 to 22 (the pulse width), and then 0 from tick 22 to 586 (the clock period), and then it repeats.
  \item \this is a signal which denotes the current mega-cell value $\mathit{GOL}(z,0)$ for 586 ticks, then $\mathit{GOL}(z,1)$ (the next time step), and so on.
\end{itemize}

For example, the wire at the top of the latch in Figure~\ref{fig:mega-cell} has the value $\this^{[-15]}$, which means that in the initial state it is holding the current value of this cell, and it will continue to hold that value until tick $586-15$, at which point it will switch to holding the next value that this cell should have, and it will switch again 586 ticks later.

Approximate signals allow arbitrary boolean combinations of the variables $\cell(m,n)$, which denote the value of a nearby cell $\mathit{GOL}(x+m,y+n,t)$. Note that $\this^{[m]}\subseteq \cell(0,0)^{[n]}$ provided that $0\le m \le n$, and we perform this ``decay'' operation early so that most of the gates never have to see an exact signal.

The value $\top$ represents failure and any signal satisfies it; it is used whenever operations go outside the expected bounds. For example, when applying the negation function to \this, since there is no \HOLTyOp{evalue} representing $\neg\this$, the result is instead $\top$.

The upshot of this is that we now get much nicer provable gate specifications:
\begin{holthmenv}
  \HOLTokenTurnstile{}\;\HOLConstNL{circuit_run'}\ \{(0,0),(2,0),(0,2),(2,2)\}\ \{((-1,0),\HOLConst{E},{\color{red}A}),((-1,2),\HOLConst{E},{\color{blue}B})\}\\
~~~~~\{((3,0),\HOLConst{E},{\color{red}A}^{[15]}\oplus{\color{blue}B}^{[18]}),((3,2),\HOLConst{E},{\color{red}A}^{[17]}\wedge{\color{blue}B}^{[15]}\}\\
~~~~~\HOLConstNL{half_adder_gate_pattern}
\end{holthmenv}

\subsection{Building the mega-cell}\label{sec:floodfill}

For the main part of the mega-cell construction, we build up a set of gates, respecting all the inputs and output relations. For this, we will make use of the \HOLConst{floodfill} function.

The basic idea is that we will build up a state consisting of the following components:

\begin{itemize}
  \item $\HOLFreeVar{area}\subseteq 2\mathbb{Z}\times2\mathbb{Z}$ is a set of points which currently contain a gate or part of a gate. This is to ensure new gates do not overlap existing gates. This set must also stay contained within $[0,42)^2$ which is a single mega-cell (the extent of figure~\ref{fig:mega-cell}).
  \item $\HOLFreeVar{ins}:(\mathbb{Z}^2\times\HOLTyOp{dir}\times\HOLTyOp{value})\;\HOLTyOpNL{list}$ is the list of unmatched input ports, and the values that they carry (treated up to permutation).
  \item $\HOLFreeVar{outs}:(\mathbb{Z}^2\times\HOLTyOp{dir}\times\HOLTyOp{value})\;\HOLTyOpNL{list}$ is the list of unmatched output ports (up to permutation).
  \item $\HOLFreeVar{crosses}:(\mathbb{Z}^2\times\mathbb{Z}^2\times\HOLTyOp{dir})\;\HOLTyOpNL{list}$ is the list of unmatched crossovers (up to permutation), which are treated specially because they do not yet have values associated with them.
  \item $\HOLFreeVar{gates}:(\mathbb{Z}^2\times\HOLTyOp{gate})\;\HOLTyOpNL{list}$ is the resulting gate list.
\end{itemize}

The definition of $\HOLConst{floodfill}\ \HOLFreeVar{area}\ \HOLFreeVar{ins}\ \HOLFreeVar{outs}\ \HOLFreeVar{crosses}\ \HOLFreeVar{gates}$ is somewhat complex, and it is easier to understand it in terms of the theorems it satisfies (simplified for presentation).

\begin{itemize}
\item \HOLThm{floodfill_add_ins}: Disjoint gate insertion.\footnote{The expression $\HOLFreeVar{area}_1+p$ is used to mean translating the set $\HOLFreeVar{area}_1$ by $p:\mathbb{Z}^2$. Similarly, $\HOLFreeVar{ins}_1+p$ translates the port positions by $p$.} The side conditions assert that $\HOLFreeVar{a}_1+p$ is disjoint from $\HOLFreeVar{a}$ and contained in $[0,42)^2$. Furthermore, the inputs in $\HOLFreeVar{i}_1$ must be exact and must not collide with any other inputs or outputs.
  \begin{holthmenv}
    \HOLConst{floodfill}\ \HOLFreeVar{a}\ \HOLFreeVar{i}\ \HOLFreeVar{o}\ \HOLFreeVar{c}\ \HOLFreeVar{gs},\quad \HOLConst{is_gate}\ g\ \HOLFreeVar{a}_1\ \HOLFreeVar{i}_1\ \HOLFreeVar{o}_1,\quad (\forall (\HOLTokenUnderscore,\HOLTokenUnderscore,v)\in i_1.\ \HOLConstNL{is_exact}\;\HOLFreeVar{v}),\\
  ~(\HOLFreeVar{a}_1+p)\cap \HOLFreeVar{a}=\emptyset,\quad (\HOLFreeVar{a}_1+p)\subseteq [0,42)^2,\quad (\HOLFreeVar{i}_1+p)\cap \HOLConstNL{fst}[\HOLFreeVar{i}\HOLTokenDoublePlus\HOLFreeVar{o}]=\emptyset\\
  \HOLTokenTurnstile{}\;\HOLConst{floodfill}\ ((\HOLFreeVar{a}_1+p)\HOLTokenDoublePlus\HOLFreeVar{a})\ ((\HOLFreeVar{i}_1+p)\HOLTokenDoublePlus\HOLFreeVar{i})\ ((\HOLFreeVar{o}_1+p)\HOLTokenDoublePlus\HOLFreeVar{o})\ \HOLFreeVar{c}\ ((p,g)::\HOLFreeVar{gs})
  \end{holthmenv}
\item \HOLThm{floodfill_add_gate}: Adding a regular gate. The inputs to the gate must match pre-existing outputs, which are removed from the list.
  \begin{holthmenv}
    \HOLConst{floodfill}\ \HOLFreeVar{a}\ \HOLFreeVar{i}\ ((\HOLFreeVar{i}_1+p)\HOLTokenDoublePlus\HOLFreeVar{o})\ \HOLFreeVar{c}\ \HOLFreeVar{gs},\quad \HOLConst{is_gate}\ g\ \HOLFreeVar{a}_1\ \HOLFreeVar{i}_1\ \HOLFreeVar{o}_1,\\
  ~(\HOLFreeVar{a}_1+p)\cap \HOLFreeVar{a}=\emptyset,\quad (\HOLFreeVar{a}_1+p)\subseteq [0,42)^2\\
  \HOLTokenTurnstile{}\;\HOLConst{floodfill}\ ((\HOLFreeVar{a}_1+p)\HOLTokenDoublePlus\HOLFreeVar{a})\ \HOLFreeVar{i}\ ((\HOLFreeVar{o}_1+p)\HOLTokenDoublePlus\HOLFreeVar{o})\ \HOLFreeVar{c}\ ((p,g)::\HOLFreeVar{gs})
  \end{holthmenv}
\item \HOLThm[floodfill_add_crossover_l]{floodfill_add_crossover}. Matches one input and puts the other one on the work queue.
  \begin{holthmenv}
    \HOLConst{floodfill}\ \HOLFreeVar{a}\ \HOLFreeVar{i}\ (({\color{red}\HOLFreeVar{ia}}+p,{\color{red}A})::\HOLFreeVar{o})\ \HOLFreeVar{c}\ \HOLFreeVar{gs},\quad \HOLConstNL{is_crossover}\ g\ \HOLFreeVar{a}_1\ [{\color{red}\HOLFreeVar{ia}},{\color{blue}\HOLFreeVar{ib}}]\ [{\color{red}\HOLFreeVar{oa}},{\color{blue}\HOLFreeVar{ob}}],\\
  ~(\HOLFreeVar{a}_1+p)\cap \HOLFreeVar{a}=\emptyset,\quad (\HOLFreeVar{a}_1+p)\subseteq [0,42)^2\\
  \HOLTokenTurnstile{}\;\HOLConst{floodfill}\ ((\HOLFreeVar{a}_1+p)\HOLTokenDoublePlus\HOLFreeVar{a})\ \HOLFreeVar{i}\ (({\color{red}\HOLFreeVar{oa}}+p,{\color{red}A}^{[5]})::\HOLFreeVar{o})\ (({\color{blue}\HOLFreeVar{ib}}+p,{\color{blue}\HOLFreeVar{ob}}+p)::\HOLFreeVar{c})\ ((p,g)::\HOLFreeVar{gs})
  \end{holthmenv}
\item \HOLThm{floodfill_finish_crossover}. Matches its second input with a pre-existing output.
  \begin{holthmenv}
    \HOLConst{floodfill}\ \HOLFreeVar{a}\ \HOLFreeVar{i}\ (({\color{blue}\HOLFreeVar{ib}},{\color{blue}B})::\HOLFreeVar{o})\ (({\color{blue}\HOLFreeVar{ib}},{\color{blue}\HOLFreeVar{ob}})::\HOLFreeVar{c})\ \HOLFreeVar{gs}\\
  \HOLTokenTurnstile{}\;\HOLConst{floodfill}\ \HOLFreeVar{a}\ \HOLFreeVar{i}\ (({\color{blue}\HOLFreeVar{ob}},{\color{blue}B}^{[5]})::\HOLFreeVar{o})\ \HOLFreeVar{c}\ \HOLFreeVar{gs}
  \end{holthmenv}
\item \HOLThm{floodfill_teleport}: ``Teleports'' an output by a multiple of 42 half-tiles. This is how the mega-cell communicates with neighboring mega-cells. Here the operation $A+z$ for $A:\HOLTyOp{value}$, $z:\mathbb{Z}^2$ does $\cell\;p+z=\cell\;(p+z)$ and distributes over boolean operations.
  \begin{holthmenv}
    \HOLConst{floodfill}\ \HOLFreeVar{a}\ \HOLFreeVar{i}\ ((p,A)::\HOLFreeVar{o})\ \HOLFreeVar{c}\ \HOLFreeVar{gs}\\
  \HOLTokenTurnstile{}\;\HOLConst{floodfill}\ \HOLFreeVar{a}\ \HOLFreeVar{i}\ ((p+42z,A+z)::\HOLFreeVar{o})\ \HOLFreeVar{c}\ \HOLFreeVar{gs}
  \end{holthmenv}
\end{itemize}

These lemmas enable the circuit to be built up by starting from the empty set, disjointly inserting a gate at the clock and at the latch, then using \HOLThm{floodfill_add_gate} to add the rest of the gates, in propagation order. The crossover handling is because \HOLThm{floodfill_add_gate} requires all inputs to a gate to have their values assigned before the gate can be added, but a quick glance at figure~\ref{fig:mega-cell} confirms that the wire leading out of the latch doesn't get very far before having to duck under a crossing wire which has not yet received a value. So in these cases we put it on the crossover queue with \HOLThm[floodfill_add_crossover_l]{floodfill_add_crossover}, and pop it off with \HOLThm{floodfill_finish_crossover} when we get the value for the other wire.

We will discuss $\HOLConst{is_gate}\ g\ \HOLFreeVar{a}_1\ \HOLFreeVar{i}_1\ \HOLFreeVar{o}_1$ further in section~\ref{sec:gate-sim}, but this specification is similar to \HOLConstNL{circuit_run'} from section~\ref{sec:approximate}, and in particular it is proved with variables for the input values, so when we use \HOLThm{floodfill_add_gate} repeatedly we build up large expressions on the outputs representing the resulting values.

\subsection{Satisfying the floodfill lemmas}\label{sec:floodfill-proof}

The advantage of the lemmas in section~\ref{sec:floodfill} is that they are easily computable --- the expressions $\HOLFreeVar{area},\HOLFreeVar{ins},\HOLFreeVar{outs},\HOLFreeVar{crosses},\HOLFreeVar{gates}$ are all concrete expressions, which makes it easy to compose the lemmas by applying them and simplifying the results. However, they entail a rather sophisticated logical structure for \HOLConst{floodfill} itself. So in this section we instead look at the definition of \HOLConst{floodfill}, and how it connects to the circuit specifications from section~\ref{sec:circuit-run}.

For $s:\HOLTyOp{stream}$ and $v:\HOLTyOp{value}$, let $s\vdash v$ mean that
$s$ is in the denotation of $v$ as described in section~\ref{sec:approximate}. Now $\HOLConst{floodfill}\ \HOLFreeVar{area}\ \HOLFreeVar{ins}\ \HOLFreeVar{outs}\ \HOLFreeVar{crosses}\ \HOLFreeVar{gates}$ holds if:

\begin{itemize}
  \item for all $(p,g)\in \HOLFreeVar{gates}$, $g$ is valid to be placed at $p$; and
  \item there exist $s_\mathrm{in}:\HOLTyOp{stream}\;\HOLTyOpNL{list}$ and $s_\mathrm{out}:\HOLTyOp{stream}\;\HOLTyOpNL{list}$ such that:
  \begin{itemize}
  \item $\forall i.\;(s_\mathrm{in})_i\vdash \HOLFreeVar{ins}_i$ and $\forall i.\;(s_\mathrm{out})_i\vdash \HOLFreeVar{outs}_i$, and
  \item for all $s_\mathrm{cross}:\HOLTyOp{stream}\;\HOLTyOpNL{list}$ such that $|s_\mathrm{cross}|=|\HOLFreeVar{crosses}|$ and $\forall i.\;\exists v.\;(s_\mathrm{cross})_i\vdash v$,\\
    \HOLConst{floodfill_run}\;\HOLFreeVar{ins'}\;\HOLFreeVar{outs'} holds, where
    \begin{itemize}
    \item $\HOLFreeVar{ins'}\;\HOLTokenDefEquality\;\overline{(\HOLConstNL{fst}\;\HOLFreeVar{ins}_i,(s_\mathrm{in})_i)}\HOLTokenDoublePlus\overline{(\HOLConstNL{in}\;\HOLFreeVar{crosses}_i,(s_\mathrm{cross})_i)})$
    \item $\HOLFreeVar{outs'}\;\HOLTokenDefEquality\;\overline{(\HOLConstNL{fst}\;\HOLFreeVar{out}_i,(s_\mathrm{out})_i)}\HOLTokenDoublePlus\overline{(\HOLConstNL{out}\;\HOLFreeVar{crosses}_i,(s_\mathrm{cross})_i)}$.
    \end{itemize}
  \end{itemize}
\end{itemize}

This part of the definition handles the delayed assignments to crossovers, and the conversion from \HOLTyOp{value} assignments to  \HOLTyOp{stream} assignments. At the core of it is another definition $\HOLConst{floodfill_run}\;\HOLFreeVar{ins}\;\HOLFreeVar{outs}$, where $\HOLFreeVar{ins},\HOLFreeVar{outs}:(\mathbb{Z}^2\times\HOLTyOp{dir}\times\HOLTyOp{stream})\;\HOLTyOpNL{list}$.
Let $S^*=\{p+42z\mid p\in S,z\in\mathbb{Z}^2\}$. Then $\HOLConst{floodfill_run}\;\HOLFreeVar{ins}\;\HOLFreeVar{outs}$ holds if:

\begin{itemize}
  \item $\HOLFreeVar{area}\subseteq [0,42)^2\cap (2\mathbb{Z}\times2\mathbb{Z})$;
  \item for all $(p,d,\HOLTokenUnderscore)\in \HOLFreeVar{ins}$, $p+d\in \HOLFreeVar{area}$;\hfill\puteqnum\label{eqn:floodfill-run-1}
  \item for all $(p,d,\HOLTokenUnderscore)\in \HOLFreeVar{outs}$, $p-d\in \HOLFreeVar{area}$;\hfill\puteqnum\label{eqn:floodfill-run-2}
  \item $\HOLConstNL{map}\;f\;\HOLFreeVar{ins}$ and $\HOLConstNL{map}\;f\;\HOLFreeVar{outs}$ have no duplicates, where $f(p,d,\HOLTokenUnderscore)=(\{p\}^*,d)$; and
  \item if
  \begin{itemize}
    \item for all $(p,d,v)\in \HOLFreeVar{ins}$, if $p-d\in \HOLFreeVar{area}^*$ then $\exists z.\;(p+42z,d,v+z)\in \HOLFreeVar{outs}$\hfill\puteqnum\label{eqn:floodfill-io-wf-1}
    \item for all $(p,d,v)\in \HOLFreeVar{outs}$, if $p+d\in \HOLFreeVar{area}^*$ then $\exists z.\;(p+42z,d,v+z)\in \HOLFreeVar{ins}$\hfill\puteqnum\label{eqn:floodfill-io-wf-2}
  \end{itemize}
  then $\HOLConst{circuit_run}\;\HOLFreeVar{area}^*\;(f\;\HOLFreeVar{ins})\;(f\;\HOLFreeVar{outs})$, where
  \begin{itemize}
    \item $f\;\HOLFreeVar{ls}\;\HOLTokenDefEquality\;\{(p+42z,d,v\;z)\mid (p,d,v)\in \HOLFreeVar{ls}, z\in\mathbb{Z}^2\}$.
  \end{itemize}
\end{itemize}

Needless to say, this definition was tricky to discover, and much of the hard work of the formalization was spent showing that all of the lemmas of section~\ref{sec:floodfill} hold for this definition.

One interesting part of the definition above is the precondition (\ref{eqn:floodfill-io-wf-1},\ref{eqn:floodfill-io-wf-2}) before \HOLConst{circuit_run}, which means that it is possible for \HOLConst{floodfill} to allow adding a gate even if it has an output which points at a gate with no matching input. Intuitively, this is okay because the \HOLFreeVar{outs} in \HOLConst{floodfill} are effectively proof obligations, and they cannot be discharged without adding another gate on the other side of the output to match it.

Another important observation is that the \HOLConst{circuit_run} expression is over $\HOLFreeVar{area}^*$, so it is an infinite circuit, tiled over space. When we add a gate, we first use the infinitary version of the composition lemma from section~\ref{sec:circuit-comp} to duplicate the gate over all of space, then use the binary composition lemma to combine that with the current floodfill area. The preconditions (\ref{eqn:floodfill-run-1},\ref{eqn:floodfill-run-2},\ref{eqn:floodfill-io-wf-1},\ref{eqn:floodfill-io-wf-2}) are used to prove the various side conditions in the composition lemma.

\section{GOL Circuit Proofs via Computation}
\label{sec:automation}

\subsection{Symbolic evaluation of individual gates}
\label{sec:gate-sim}

We prove individual gate specifications (in terms of \HOLConst{circuit_run} from Section~\ref{sec:circuit-run})
by symbolically simulating the steps a circuit and its IO interfaces cycle through (Section~\ref{sec:circuitio}).
That is, rather than simulating concrete GOL states, in our simulator each GOL cell contains a \HOLConst{bexp}, given by the following grammar:
\vspace{-7pt}

$$\HOLTyOp{bexp}\;=\;\top\mid \bot\mid \HOLTyOp{var}_\mathbb{N}\mid \neg\HOLTyOp{bexp}\mid \HOLTyOp{bexp}\wedge\HOLTyOp{bexp}\mid \HOLTyOp{bexp}\vee\HOLTyOp{bexp},\qquad
  \HOLTyOp{var}\;=\;{\color{red}\HOLConst{A}}\mid {\color{blue}\HOLConst{B}}$$
Our symbolic simulations are set up to take two input streams, {\color{red}\HOLConst{A}} and {\color{blue}\HOLConst{B}}. Variable ${\color{red}\HOLConst{A}}_3$ is the value at index 3 of the input stream {\color{red}\HOLConst{A}}. (We call this index the ``age'' of the variable.)

The goal of the symbolic simulations is to prove circuit specifications for individual gates that describe all the states a gate can be in, including all the states the internal in-progress LWSSs can be in.
We do this in two steps:
\begin{enumerate}
\item Outside of the proof assistant, we run several ticks of symbolic simulation starting from a completely concrete initial state but feeding in variables in the places where input LWSSs are to arrive. We are done once these LWSSs have propagated through the entire gate.
\item Inside the proof assistant, we run one tick of symbolic simulation according to section~\ref{sec:circuitio} on the symbolic state found outside of the proof assistant. At the end of the in-prover simulation, we check that the resulting symbolic state is equal to the initial symbolic state but with every variable aged by one, e.g., ${\color{red}\HOLConst{A}}_3$ becomes ${\color{red}\HOLConst{A}}_4$.
\end{enumerate}
The final check shows that gates have stable behaviour over any number of ticks and function like LWSS conveyor belts, with each LWSS making steady progress through the gate.

Figure~\ref{fig:and} shows the verified symbolic state of the AND gate, using color to indicate the cells with variable expressions. (Age is not represented in the diagram, but each LWSS in the diagram has a different age. The leftmost LWSS is ${\color{red}\HOLConst{A}}_4$, and the rightmost is ${\color{red}\HOLConst{A}}_0\wedge{\color{blue}\HOLConst{B}}_0$.)

The most interesting aspect of the symbolic simulation is how we determine the next symbolic state of a GOL cell given the current symbolic state of it and its neighbors.
The approach we take is to:
\begin{enumerate}
\item collect all the variables appearing in any of the relevant symbolic cells,
\item for each possible assignment to those variables, we evaluate the GOL rules,
\item we summarize all results in one expression as a nested $\HOLKeyword{if}$-expression,
\item we simplify the $\HOLKeyword{if}$-expressions and represent it as a single \HOLConst{bexp}.
\end{enumerate}
Step 3 can, for example, result in the following $\HOLKeyword{if}$-expression
if a GOL cell and its neighbors use two variables, ${\color{red}\HOLConst{A}}_2$ and ${\color{blue}\HOLConst{B}}_1$, and only evaluates
to true if both are true.
\[\HOLKeyword{if}~{\color{red}\HOLConst{A}}_2~\HOLKeyword{then}~(\HOLKeyword{if}~{\color{blue}\HOLConst{B}}_1~\HOLKeyword{then}~\textsf{true}~\HOLKeyword{else}~\textsf{false})~\HOLKeyword{else}~\textsf{false} \]
The expression above simplifies to ${\color{red}\HOLConst{A}}_2\,{\land}\,{\color{blue}\HOLConst{B}}_1$.
This approach would be unworkable if the number of variables exceeds 2 or 3.
Fortunately, such situations are very rare in our gate simulations (it goes as high as 4 variables in the half-adder, see section~\ref{sec:approximate}).

\subsection{Putting it all together}

The \HOLConst{floodfill} lemmas of section~\ref{sec:floodfill} are designed so that the main part of the construction can be done fully automatically.
\begin{itemize}
\item An SML program takes as input an ASCII-art version of Figure~\ref{fig:mega-cell}.
\item It is parsed to get gate positioning and orientation, and the appropriate symbolic evaluation theorem from section~\ref{sec:gate-sim} is selected.
\item The initial gates (at the latch and clock), with their values, are additional inputs, and it uses \HOLThm{floodfill_add_ins} to add these.
\item It then performs a depth first traversal of the diagram.
\begin{itemize}
  \item If an output is facing a gate which has all of its inputs in the output list, use ${\HOLThm{floodfill_add_gate}}$.
  \item If an output is facing a crossover, use \HOLThm[floodfill_add_crossover_l]{floodfill_add_crossover}, or \HOLThm{floodfill_finish_crossover} if it is the second time we have visited this gate.
  \item If an output is facing the edge of the tile, use \HOLThm{floodfill_teleport} to wrap it back in bounds.
\end{itemize}
These steps are repeated until nothing can make progress. In the process we work out all of the formulas associated to each IO port.
\item The \HOLTyOp{value} type from section~\ref{sec:approximate} has functions defined on it for $\wedge$, $\vee$, $\neg$, which do the obvious thing on \HOLTyOp{avalue} values, but there are a few interesting cases designed to handle the latch area:
\begin{itemize}
  \item $\this^{[m]}\wedge (\neg \ck)^{[n]}=(\this\wedge \neg \ck)^{[n]}$ provided $n\le m\le n+22$
  \item $\ck^{[m]}\wedge v^{[n]}=(\this\wedge \ck)^{[m]}$ provided $n\le m+586$, $m\le -22$, and $v=\HOLConst{nextCell}$
  \item $(\this\wedge \ck)^{[n]}\lor (\this\wedge \neg\ck)^{[n]}=\this^{[n]}$
\end{itemize}
Here $\HOLConst{nextCell}:\HOLTyOp{avalue}$ is the specific formula that the mega-cell circuit computes.
\item $\HOLConst{nextCell}$ is a boolean combination of $\cell(m,n)$ values for $-1\le m,n\le 1$, and we prove it is equal to the GOL step function from section~\ref{sec:gol-rules} by enumerating the 512 possibilities.
\item The final $\HOLConst{floodfill}$ theorem has only two outputs, which overlap the two inputs, and therefore they cancel and produce a complete GOL (not GOL-IO) simulation. In particular, since one of these inputs has value $\this^{[-15]}$, we know that in the final simulation, if we sample a particular pixel in this IO port at multiples of 586 ticks, it will be on iff the corresponding mega GOL simulation pixel is on.
\end{itemize}

\noindent In the end, the final theorem we obtain looks like this (\HOLThm{gol_in_gol_circuit_thm}):
\begin{holthmenv}
  \HOLTokenTurnstile{}\;\forall\,n\;S.\;\HOLConst{step}^{n}\;S =\HOLConst{read_mega_cells}\;(\HOLConst{step}^{n \times 60 \times 586}\;(\HOLConst{build_mega_cells}\;S))
\end{holthmenv}
where $\HOLConst{step}$ is the GOL step function; $\HOLConst{build_mega_cells}\;s$ takes an input GOL state $S$ and tiles the plane with two versions of the mega-cell in figure~\ref{fig:mega-cell}, which differ only slightly, in the internal state of the latch; and $\HOLConst{read_mega_cells}\;S=\{p\mid 3150p+(1726, 599)\in S\}$ performs the aforementioned sampling.

\section{Conclusion, Related Work and Future Work}
\label{sec:conclusions}

In this paper, we have demonstrated that it is possible to formally verify circuits built in GOL and we have verified a circuit that implements GOL itself inside GOL. To the best of our knowledge, this is the first work to formally verify, in an interactive theorem prover (ITP), constructions in a cellular automata. The formalization is roughly 9\,500 LOC.

There has been significant prior work on formalizing more traditional models of computation in ITPs, e.g., Turing machines \cite{ForsterTuring,AspertiR12,XuZU13,Ciaffaglione16}, register machines \cite{ForsterL19,BayerDPSS19}, $\lambda$-calculus \cite{Norrish11,ForsterS19,ForsterSmolka,ForsterKR20}, $\mu$-recursive functions \cite{Carneiro18} and more~\cite{RamosMAMDN18}. We refer to Forster~\cite{Forster21} for more in depth discussions on computability in ITPs. Rule 110 \cite{rule110} is another simple universal CA.

In future work, it would be interesting to explore ITP proofs connecting GOL with more traditional forms of computability. Also, the tools used here could be generalized to prove other GOL circuits, other cellular automata, as well as low level hardware correctness proofs.

%
\bibliography{paper}

\end{document}